\documentclass[aps,prc,twocolumn,superscriptaddress,showpacs,floatfix]{revtex4}
\usepackage{graphicx,color}
\usepackage{braket}
\usepackage{amsmath}
\usepackage{amssymb}
\newlength{\figwidth}
\def\scr{\rm\scriptscriptstyle }

\begin{document}
\setlength{\figwidth}{0.98\columnwidth}

\title[]{Theoretical study of the elastic breakup of weakly bound nuclei at near barrier energies}
\author{D.R. Otomar,$^1$ P. R. S. Gomes,$^1$ J. Lubian}
\affiliation{Instituto de F\'{\i}sica, Universidade Federal Fluminense, Av. Litoranea
s/n, Gragoat\'{a}, Niter\'{o}i, R.J., 24210-340, Brazil}
\author{L. F. Canto}
\affiliation{Instituto de F\'{\i}sica, Universidade Federal Fluminense, Av. Litoranea
s/n, Gragoat\'{a}, Niter\'{o}i, R.J., 24210-340, Brazil and Instituto de F%
\'{\i}sica, Universidade Federal do Rio de Janeiro, CP 68528, Rio de
Janeiro, Brazil}
\author{M. S. Hussein}
\affiliation{Instituto de Estudos Avan\c{c}ados, Universidade de S\~{a}o Paulo C. P.
72012, 05508-970 S\~{a}o Paulo-SP, Brazil, Instituto de F\'{\i}sica,
Universidade de S\~{a}o Paulo, C. P. 66318, 05314-970 S\~{a}o Paulo,-SP,
Brazil, and  Departamento de F\'{\i}sica, Instituto Tecnol\'{o}gico de Aeron\'{a}utica, DCTA,12.228-900 S\~{a}o Jos\'{e} dos Campos, SP, Brazil,}

\begin{abstract}
We have performed CDCC calculations for collisions of $^{7}$Li projectiles on  $^{59}$Co, $^{144}$Sm and $^{208}$Pb targets
at near-barrier energies, to assess the importance of the Coulomb and the nuclear couplings in the breakup of $^{7}$Li, as well 
as the Coulomb-nuclear interference.  We have also investigated scaling laws, expressing the dependence of the cross sections 
on the charge and the mass of the target. This work is complementary to the one previously reported by us on the breakup of
$^{6}$Li. Here we explore the similarities and differences between the results for the two Lithium isotopes. The 
relevance of the Coulomb dipole strength at low energy for the two-cluster projectile is investigated in details.
\end{abstract}

\maketitle

\section{Introduction}

Reaction mechanisms in collisions of weakly bound nuclei have been intensively investigated in the last 
years~\cite{CGD06,LiS05,KRA07,KAK09,HaT12,BEJ14,CGD15}, both theoretically and experimentally. These mechanisms may 
be particularly interesting in collisions of halo nuclei, where the breakup process and its influence on other reaction channels,
such as fusion, tend to be very strong. 
However, the processes involved in collisions of stable weakly bound nuclei, like $^{6}$Li, $^{7}$Li and $^{9}$Be, are expected to
be qualitatively similar. On the other hand, the intensities of stable beams are several orders of magnitude larger than those presently 
available  for radioactive beams. For this reason, collisions of stable weakly bound nuclei have been widely studied. Since 
performing direct measurements of breakup cross sections is a very hard task, most experiments determine fusion and elastic
cross sections. Recent experiments have shown that transfer processes followed by breakup may predominate over direct
breakup of stable weakly bound nuclei at sub-barrier energies \cite{Luong,Dasgupta10,Rafiei, Shiravasta}. \\

In a recent paper \cite{Otomar13} we have reported continuum discretized coupled channel (CDCC) calculations for collisions 
of $^6$Li projectiles with $^{59}$Co, $^{144}$Sm and $^{208}$Pb targets at near-barrier energies. We have evaluated  Coulomb, 
nuclear and total breakup angular distributions, as well as the corresponding integrated cross sections. We have observed strong
Coulomb-nuclear interference, and found that the nuclear and the Coulomb components of the breakup cross sections follow 
scaling laws. For the same $E/V_{\scr B}$ (energy normalized to the Coulomb barrier),  the nuclear component of
the breakup cross section is proportional to $A_{\scr T}^{\scr 1/3}$, where $A_{\scr T}$ is the target's mass number.
An explanation for this behavior was latter given by Hussein \textit{et al.} \cite{Hussein13}. On the other hand, the Coulomb breakup
component was shown to depend linearly on the target's atomic number, $Z_{\scr T}$. In the present paper we complement the 
previous work by performing the same kind of analysis for  $^{7}$Li projectiles.\\

There are two important differences between the $^{6}$Li and $^{7}$Li Lithium isotopes. The first is that the breakup threshold 
energy, or Q-value, of $^{7}$Li  is about 1 MeV lower than that of $^{7}$Li. They are respectively 1.47 and 2.47 MeV. The second 
difference is that $^{7}$Li has a non-zero low energy dipole strength, contrary to $^{6}$Li. Their dipole responses are related to their 
cluster structure ($\alpha$ - t and $\alpha$ - d for the $^7$Li and $^6$Li, respectively). In fact, using the cluster model, the $B(E1)$ distribution  in the 
projectile $a = c + p$, is given by \cite{hus2,hus3},
\begin{multline}
\frac{dB(E1)}{dE_x} = S\,N_{0} ^{2}\, \frac{3}{\pi^2}\ \left( \frac{\hbar^2}{\mu_{cp}} \right)^{2} \frac{Q^{1/2}\,(E_x - Q)^{3/2}}{E_{x}^{4}} 
\\
 \times \left[\frac{Z_{p}A_{c} - Z_{c}A_{p}}{A_{a}}\right]^{2} e^2.
 \label{BE1-a}
\end{multline}
Above, $\mu_{cp}$ is the reduced mass of the $a = c + p$ system, $S$ is cluster spectroscopic factor and $N_0$ is a normalization constant which takes into account the finite range of the $c + p$ potential.
The $B(E1)$ value is obtained by integrating the above over 
$E_x$,
\begin{equation}
B(E1) = \frac{9\hbar^2}{16\pi}\frac{A_{a}}{A_{p}{A_c}}\left[\frac{Z_{p}A_{c}-Z_{c}A_{p})}{A_{a}}\right]^2 e^2 \frac{1}{Q}.
\label{BE1}
\end{equation}
Using Eq.~(\ref{BE1}), one finds  for $^7$Li: $B(E1) \simeq 10\, {\rm fm}^{2}e^2$. On the other hand, the above expression vanishes
identically  for  $^6$Li. This implies a larger Coulomb breakup for $^7$Li. In fact the Coulomb breakup of $^6$Li is dominated by
higher multipolarities, such as quadrupole.  A more detailed discussion of this issue can be found 
in \cite{hus3}.\\

As in our previous work, the choice of the $^{59}$Co, $^{144}$Sm and $^{208}$Pb targets was based on the availability of 
elastic scattering data at near-barrier energies.  In this way, we were able to check the reliability of our CDCC model 
applying it to elastic scattering and comparing the theoretical cross sections with the data.\\

The paper is organized as follows. In section II some details of our CDCC model space are given. In section III the results of our calculation are discussed, while the section IV is devoted to our conclusions.

\section{The CDCC model}

The most suitable approach to deal with the breakup process, which feeds to the population of states in the continuum, is the 
so called CDCC method~\cite{KYI86,AIK87}. In this type of calculations, the continuum wave functions are grouped into bins, or 
wave packets, that can be treated similarly to the usual bound inelastic states, since they are described by square-integrable wave
functions. In the present work we use the same assumptions and methodology of the CDCC calculations of 
Refs.~\cite{Otomar13, Otomar09,Otomar10}. We assume that $^{7}$Li breaks up directly into an $\alpha$-particle and a tritium,
with separation energy $S_{\alpha }=2.47$ MeV. To describe the breakup of the projectile into two charged fragments, we used the
cluster model. We consider that the two clusters are bound in the entrance channel and the first inelastic channel with spin $1/2^-$
and excitation energy 0.477 MeV. The remaining projectile states are all in the discretized continuum. Resonant states of the 
projectile are explicitly taken into account, to avoid double counting. In all calculations of the present work, we performed our
numerical calculations using the code FRESCO \cite{Tho88}.\\

In the standard CDCC method  \cite{KYI86,AIK87}, the scattering of a projectile, composed by a core \textit{c} (the alpha particle 
in the present work) and a valence particle \textit{p} (the triton), by a target \textit{T} is modelled by the Hamiltonian:
\begin{equation}
H=K_{rel}(\mathbf{R})+K_{int}(\mathbf{r}) + V_{pc} + U_{pT} + U_{cT},
\label{eq1}
\end{equation}
where K$_{rel}$ is the projectile-target relative kinetic motion, K$_{int}$ is projectile internal kinetic energy, V$_{pc}$ is the p-c 
binding potential and U$_{pT}$ and U$_{cT}$ are the p-T and c-T optical potentials, respectively. These optical potentials are 
chosen by the condition of describing the elastic scattering of each cluster from the target. They have an imaginary part arising
both from fusion of the cluster with the target and from the excitation of inelastic states in the target. Thus, the breakup cross 
sections obtained in standard CDCC calculations correspond only to elastic breakup. However, the influence of inelastic breakup 
on elastic scattering is taken into account through the action of the imaginary parts of $U_{pT}$ and $U_{cT}$ at the surface 
region. To go beyond the standard CDCC method, treating target excitations explicitly, one should include in Eq.~(\ref{eq1}) 
an additional term corresponding to the internal Hamiltonian of the target. This procedure is not followed in the present work, 
where only inelastic states of the projectile are included in our channel space.\\

The sum of the cluster-target potentials  of Eq.~(\ref{eq1}) gives the total interaction between the projectile and the target. It can 
be written as
\begin{equation}
U(\mathbf{R},\mathbf{r} )=U_{cT}(\mathbf{R},\mathbf{r})+U_{pT}(%
\mathbf{R},\mathbf{r}), 
\label{split}
\end{equation}
where, ${\bf R}$ is the vector joining the centers of mass of the projectile and the target, and ${\bf r}$ is the relative
position vector between the two clusters. $U(\mathbf{R},\mathbf{r} )$ gives the bare potentials 
(diagonal matrix-elements), and also all couplings among the channels (off-diagonal matrix-elements in 
channel space). This potential contains contributions from Coulomb and from nuclear forces, and the importance of each 
contribution can be assessed switching off the other.\\

Concerning the CDCC model space
for $^{7}$Li, the continuum (nonresonant and resonant) subspace is
discretized into equally spaced momentum bins with respect to the momentum $%
\hbar k$ of the $\alpha -t$ relative motion. The bin widths are suitably
modified in the presence of the resonant states in order to avoid double
counting. In this way, the discretization is as follows:  continuum partial
waves up to $l_{max}$ = 4 waves for a density of the continuum
discretization of 2 bins/MeV (l = 0,1,2); 7.7 bins/MeV and 1.92 bins/MeV
below and above the $7/2^{-}$ resonance, respectively; 10 bins/MeV
inside the resonance; 2.5 bins/MeV and 2 bins/MeV below and above the $5/2^{-}$ resonance, respectively; 2.5 bins/MeV inside the resonance; 2 bins/MeV for both $7/2^{+}$ and $9/2^{+}$ resonances. The
projectile fragments-target potential multipoles up to the term $%
K_{max}$ = 4 were considered. For the interaction $\alpha $ - tritium
to generate the bins, we use an appropriate Woods-saxon potential
to describe the unbound resonant and nonresonant states \cite%
{Otomar09,Otomar10}. For the resonant states, we included a spin-orbit
interaction. To get a finite set of coupled equations, one must truncate the discretized
continuum at some maximal value of the excitation energy and of the orbital angular momentum of clusters.
For this reason, rigorous convergence tests have to be performed. \\

\section{Numerical calculations}

We have performed CDCC calculations for the $^7$Li + $^{59}$Co, $^7$Li +$^{144}$Sm and $^7$Li +$^{208}$Pb systems,
for which elastic scattering data at near-barrier energies are available (Refs.~\cite{Beck}, \cite{Figueira10} and \cite{Keeley94},
respectively).  For the alpha-target and tritium-target optical potentials of Eq.~(\ref{split}), we used the double-folding S\~ao Paulo 
potential~\cite{Chamon97,Chamon02}. The target densities, used in the folding integrals, were taken from the systematics of 
the S\~ao Paulo potential~\cite{Chamon02}. Assuming that charge and matter densities have similar distributions, the matter 
density distribution of the triton was obtained multiplying by 3 the charge distribution reported in Ref. \cite{19}. The matter density 
of the $^4$He cluster was obtained through the same procedure. We assumed that the imaginary parts of the optical potentials have 
the same radial dependence of the real part, with a weaker strength. Then, we adopted the expression, 
$U_{jT}(r) = \left[ 1+0.78\,i\right]\,V_{\scr SPP}(r)$, with $j$ standing for either the alpha or the tritium cluster, and $V_{\scr SPP}(r)$ 
standing for the S\~ao Paulo Potential. This procedure has been able to describe the reaction cross section (and consequently the 
elastic angular distribution) for many systems in a wide energy interval \cite{gasques}. 
\begin{figure}%[h!]
\centering
%\begin {center}
\includegraphics[scale=0.40]{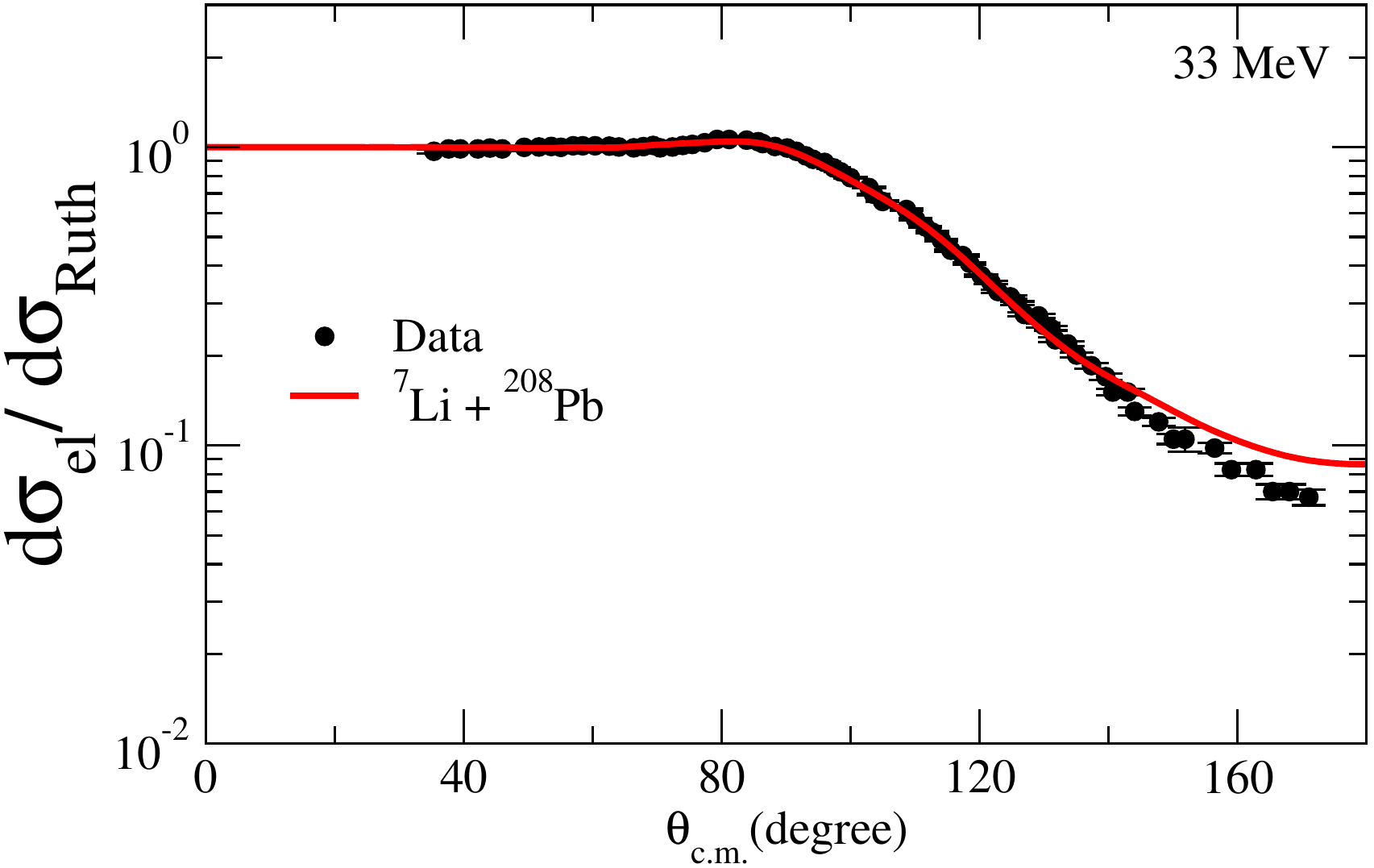}
\caption{(color on line) Comparison of elastic scattering data with predictions from our CDCC calculations for $^7$Li + $^{208}$Pb at 33.0 MeV. Data are from \cite{Keeley94}.}
\label {fig1}
%\end{center}
\end{figure}
Before calculating breakup cross sections, we made sure that our CDCC calculations were able to reproduce elastic scattering data. 
This is illustrated in Fig.~\ref{fig1}, which shows the theoretical and experimental elastic scattering cross sections for 
$^7{\rm Li} - ^{208}{\rm Pb}$ scattering at the bombarding energy $E_{\rm lab} = 33$ MeV. The agreement is good, except for
some small discrepancies at backward angles. This is quite satisfactory, if one considers that there is no adjustable parameter 
in our calculations.\\

As in Refs.~\cite{Otomar13,Hussein13}, we write the elastic breakup cross section as,
\begin{equation}
\sigma_{bup} = \sigma_{C} + \sigma_{N} + \sigma_{int}.
\end{equation}
That is, the breakup cross section is split into a Coulomb component, $\sigma_{C}$, a nuclear component, $\sigma_{N}$,
and an interference term, $\sigma_{int}$. The two components were evaluated by CDCC calculations switching off either
the nuclear or the Coulomb part of the coupling interaction. \\

To be fair, we should mention that the above procedure does not 
generate the full CDCC Coulomb and nuclear components of the cross section as these are both influenced by the each others: 
the Coulomb contribution is influenced by the nuclear scattering and the nuclear contribution is influenced by the Coulomb 
scattering. However, to perform the separation within a coupled channel framework is a very hard task. On the other hand,
this can easily be done within a Distorted Wave Born approximation (DWBA) treatment of the breakup process \cite{hus1}. 
The DWBA calculation is usually employed at higher energies or weak coupling to the breakup channel (high Q-value), and 
does not serve our purpose here. Thus we have no other choice but to use the prescription originally employed by \cite{hus4}, 
and recently used by us \cite{Otomar13,Hussein13}, of switching off the undesired interaction to obtain the desired component. 
We believe that this approximate method of generating the Coulomb and the nuclear breakup components of the coupled channels-calculated cross section
is reasonable for very light targets, such as $^{12}$C where the nuclear breakup dominates, and for very heavy targets, such as $^{208}$Pb where Coulomb breakup 
by far dominates. However, we have no way to know how accurate the switching off method in the case of medium mass targets, where both the Coulomb and nuclear
components are equally important.

It remains as an open problem the assessment of the error  inherent in such a procedure within the coupled channels theory.\\

\begin{table}
\footnotesize
\centering
\begin{tabular}{|c|c|c|c|c|}\hline
\multicolumn{5}{|c|}{$^{7}$Li $+$ $^{208}$Pb} \\
\hline
 
$E/V_{\scr B}$ & {$\sigma_{C}(mb)$} & {$\sigma_{N}(mb)$} & {$\sigma_{bup}(mb)$} & {$({\sigma_{bup} - \sigma_{N})}/ \sigma_{C}$} \\  \hline

%E/V$_{B}$ & $^{7}$Li & $^{6}$Li & $^{7}$Li & $^{6}$Li & $^{7}$Li & $^{6}$Li & $^{7}$Li & $^{6}$Li\\
%\hline

 $0,84$ & $7,28$ & $0,90$ & $4,51$& $0,50$\\ \hline
 
 $1,00$ & $11,20$  &$2,65$  & $10,31$ & $0,68$ \\ \hline
 
 $1,07$ & $16,00$  & $9,18$ & $14,94$ & $0,36$ \\ \hline
 
 $1,30$ & $31,64$  & $11,88$ & $30,48$ & $0,59$\\ \hline

\multicolumn{5}{|c|}{$^{7}$Li $+$ $^{144}$Sm} \\ \hline

 $0,84$ & $2,49$ & $0,51$ & $0,88$ & $0,15$ \\ \hline
 
 $1,00$ & $6,21$ & $2,50$ & $5,21$ & $0,44$ \\ \hline
 
 $1,07$ & $6,20$ & $6,57$ & $5,11$ & $-0,24$ \\ \hline
 
 $1,30$ & $16,09$ & $8,78$ & $18,71$ & $0,62$\\ \hline

\multicolumn{5}{|c|}{$^{7}$Li $+$ $^{59}$Co} \\ \hline

 $0,84$ & $0,17$ &  $0,05$ &  $0,23$ &  $1,06$ \\ \hline
 
 $1,00$ & $1,12$ &  $1,00$ &  $2,10$ &  $0,98$ \\ \hline
 
 $1,07$ & $1,84$ &  $2,09$ &  $3,43$ & $0,73$ \\ \hline
 
 $1,30$ & $4,34$ &  $7,08$ &  $12,04$ & $1,14$\\ \hline
 
\end{tabular}
\label{table1}
\caption{Integrated breakup cross sections for $^7$Li on $^{59}$Co, $^{144}$Sm and $^{208}$Pb targets at energies close to the Coulomb barriers. The first column correspond to the Coulomb component of the breakup, the next ones to the nuclear component and the total breakup. The last column should be equal to unity if there were no interference between the Coulomb and nuclear components. See  text for details.} 
\end{table}
Table I
%\ref{table1} 
shows the integrated $^{7}$Li breakup cross sections for the three
systems at near-barrier energies. As expected, one observes that the Coulomb and the nuclear components, as well as the total 
breakup cross sections, for the light targets are much smaller than the corresponding cross sections for the heavier targets. The
interference between the nuclear and Coulomb breakup amplitudes can be easily observed in the last column of Table I. In the
no-interference limit, the quantity $({\sigma_{bup} - \sigma_{N})}/ \sigma_{C}$ should be equal to one. The numbers shown in 
the table are very different from this limit, which indicates that there is strong Coulomb-nuclear interference in the  breakup of 
 $^7$Li. The same conclusion was reached in the case of the $^6$Li isotope~\cite{Otomar13}. \\

\begin{figure}%[h!]
\centering
%\begin {center}
\includegraphics[scale=0.45]{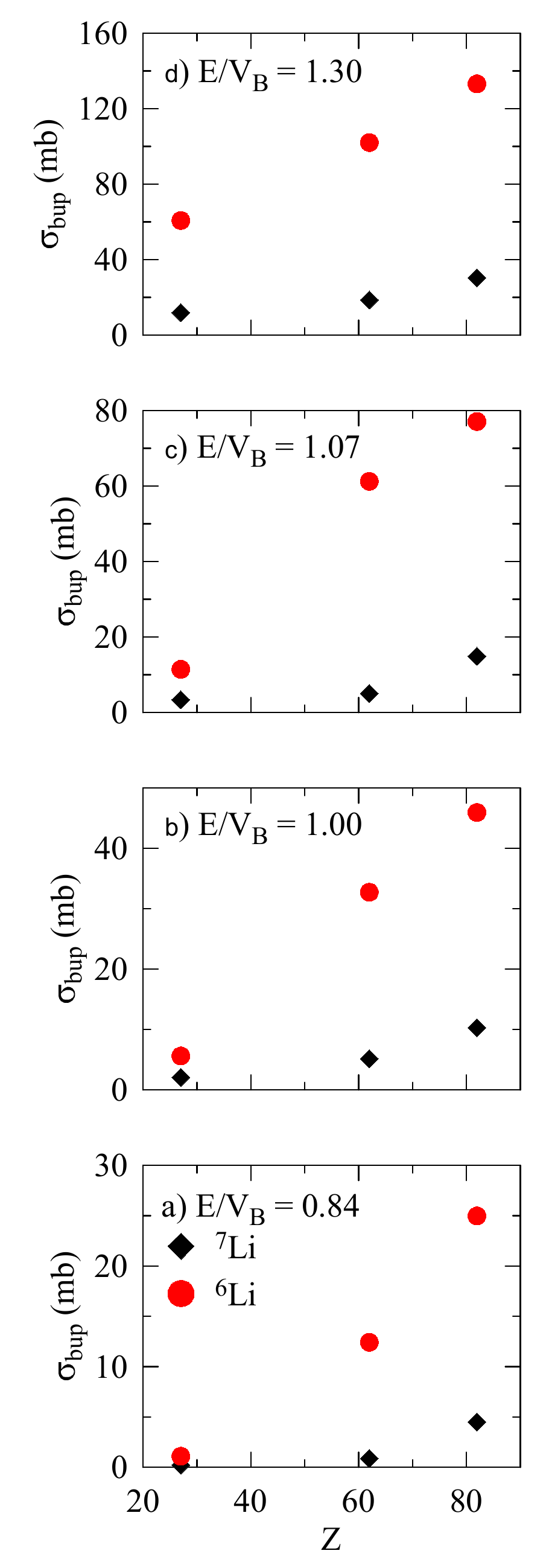}
\caption{(color on line) Total breakup cross sections for $^6$Li and $^7$Li projectiles on $^{59}$Co, $^{144}$Sm and $^{208}$Pb targets, for energies close to the Coulomb barrier. Results for $^6$Li were already published in Ref. \cite{Otomar13}.}
\label {fig2}
%\end{center}
\end{figure}

In Fig.~\ref{fig2} we show the integrated cross sections for the breakup of $^{6,7}$Li  on $^{59}$Co, $^{144}$Sm and $^{208}$Pb
targets, at three near-barrier energies. The cross sections for $^{7}$Li are results of the present calculations whereas those for
$^{6}$Li were taken from Ref.~\cite{Otomar13}.  One observes that, for a given projectile and at the same  value of $E/V_{\scr B}$, 
the breakup cross sections increases with the target charge. One sees also that, for each target and at the same relative energy, 
the cross sections for $^{6}$Li are much larger than those for $^{7}$Li. This is not surprising, since the breakup threshold energy 
for $^{6}$Li is appreciably smaller than that for $^{7}$Li. \\

\begin{figure}%[h!]
\centering
%\begin {center}
\includegraphics[scale=0.45]{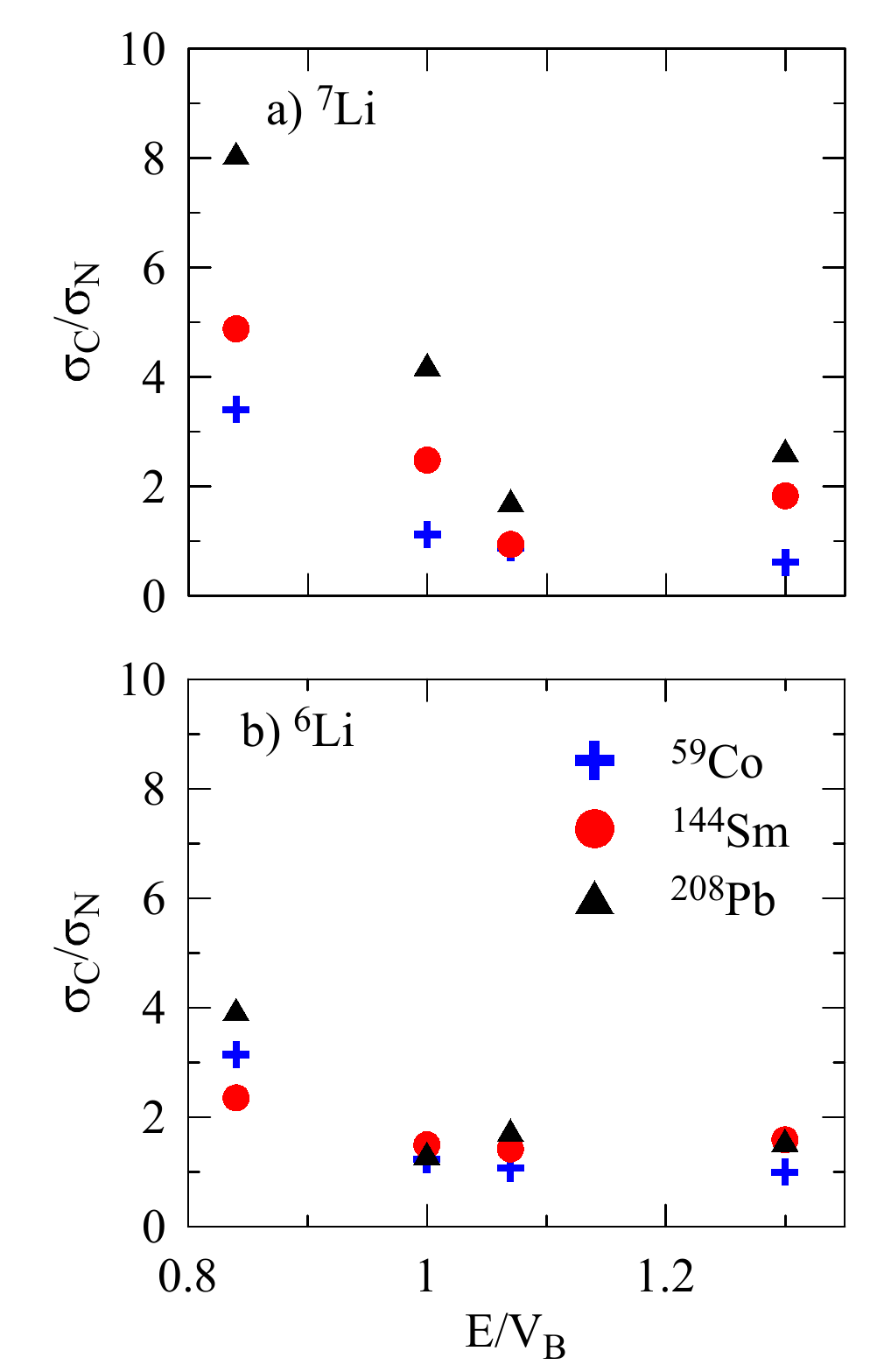}
\caption{(color on line) Ratio between Coulomb and nuclear breakup as a function of energy for the $^6$Li and $^7$Li projectiles on the $^{59}$Co, $^{144}$Sm and $^{208}$Pb targets.}
\label {fig3}
%\end{center}
\end{figure}

Using the values of the breakup cross sections given in Table I and the results of Ref.~\cite{Otomar13}, we can  plot the ratio 
$\sigma_C/\sigma_N$ as a function of the relative energy. The results for the targets considered in our study are shown in 
Fig.~\ref{fig3}, for the breakup of $^7$Li (panel a) and for the breakup of $^6$Li (panel b). One observes that this ratio 
decreases as $E/V_{\scr B}$ increases, and that it is systematically larger than one, except for the breakup of $^7$Li
on the lightest target at above-barrier energies ($E/V_{\scr B}>1$). One notices also that, for a given projectile and at
a fixed value of $E/V_{\scr B}$, the ratio increases with the charge of the target. This behavior is expected and it has already
been observed for $^{6}$Li projectiles~\cite{Otomar13}.  However, the most interesting (and new) result in Fig.~\ref{fig3} is 
that this ratio for a given target and a given $E/V_{\scr B}$ is much larger in the breakup of $^{7}$Li than in that of $^{6}$Li. 
This result should arise from the fact that the low-energy Coulomb dipole response in the breakup of $^7$Li is larger than in
the breakup of $^6$Li. The reason is that the factor $\left[  Z_{p}A_{c} - Z_{c}A_{p} \right]^2$, appearing in 
Eqs.~(\ref{BE1-a}) and (\ref{BE1}), is equal to 4 for $^7$Li, whereas in the case of $^6$Li it vanishes identically.

A detailed study of Figs.~\ref{fig2} and \ref{fig3} leads to a very interesting conclusion. The analysis of Fig.~\ref{fig2} indicated that 
the breakup cross sections for $^6$Li are larger than those for $^7$Li, even for the $^{208}$Pb target. In this case, the Coulomb
breakup dominates, as can be seen in Table I (for $^7$Li) and in Ref.~\cite{Otomar13} (for $^6$Li). However, Coulomb breakup 
depends on two factors. The first is the low-energy Coulomb dipole response, which vanishes for $^6$Li and does not for $^7$Li.
The second is the low breakup threshold, which is 1 MeV lower in the case of $^6$Li. Fig.~\ref{fig2} indicates that the predominant factor is
the lower breakup threshold of the $^6$Li projectile. On the other hand, Fig.~\ref{fig3} indicates that the ratios $\sigma _{C}$/$\sigma _{N}$  are 
systematically larger for the $^7$Li projectile. The consistency of the two above conclusions would require that the nuclear breakup of $^6$Li 
be much larger than that of $^7$Li. This can be checked comparing $\sigma _{N}$ for the two projectiles on the same target and at the same value of
E/V$_B$. Looking at the nuclear breakup cross sections in Table I  (for $^7$Li) and at those given in Ref.~\cite{Otomar13} (for $^6$Li), one 
concludes that this condition is satisfied. For example, for the $^{208}$Pb target at $E/V_{\scr B} = 0.84$, the cross sections for the nuclear breakup 
of $^6$Li and for that of $^7$Li are respectively 8.8 mb and 0.9 mb. \\

\begin{figure}%[h!]
\centering
%\begin {center}
\includegraphics[scale=0.5]{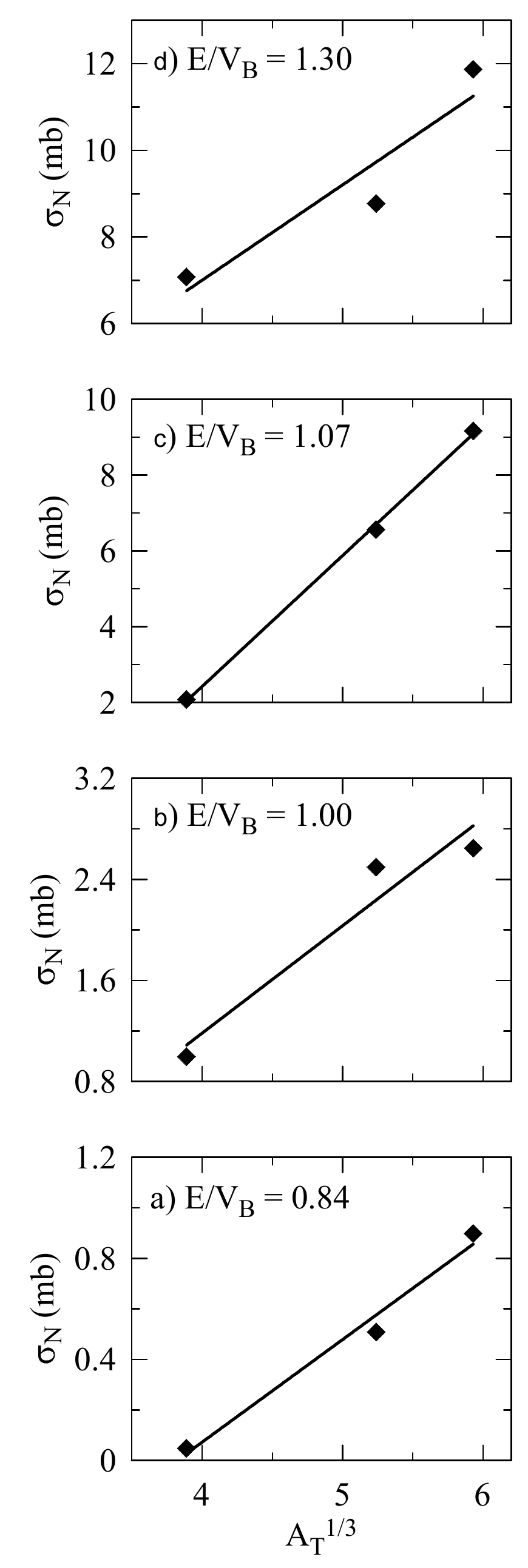}
\caption{$^7$Li nuclear breakup cross sections as a function of the target mass, for $^{59}$Co, $^{144}$Sm and $^{208}$Pb targets.}
\label {fig4}
%\end{center}
\end{figure}

\begin{figure}%[h!]
\centering
%\begin {center}
\includegraphics[scale=0.5]{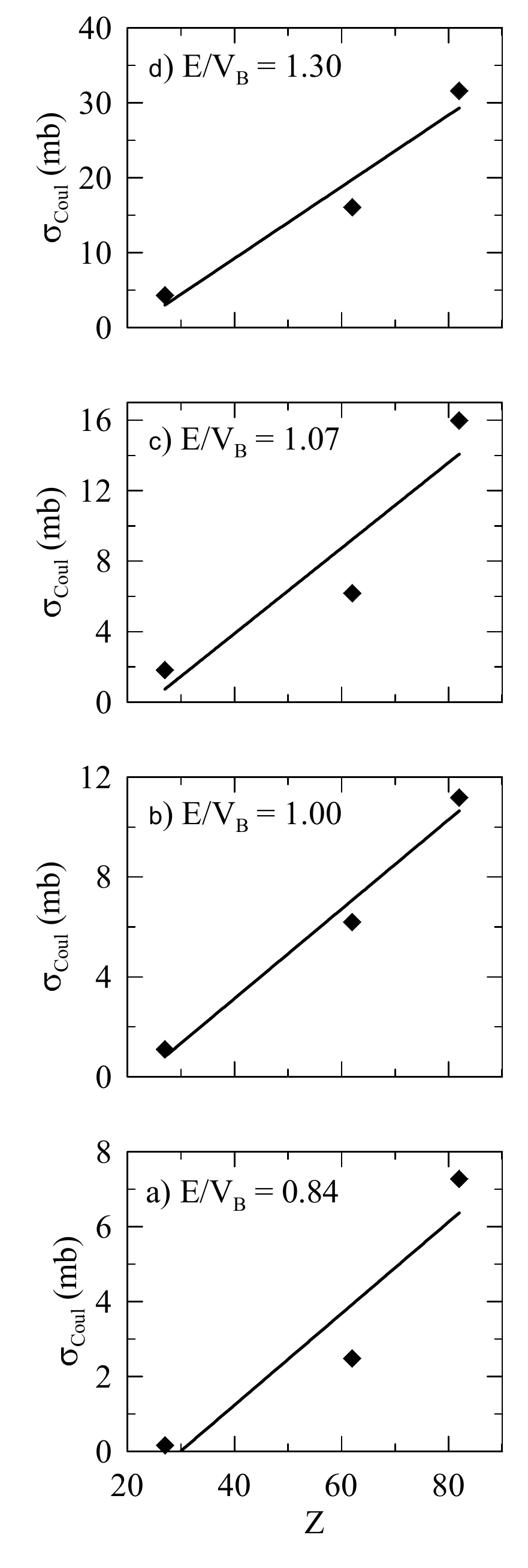}
\caption{$^7$Li Coulomb breakup cross sections as a function of the target charge, for $^{59}$Co, $^{144}$Sm and $^{208}$Pb targets.}
\label {fig5}
%\end{center}
\end{figure}

We have also investigated scaling laws in the nuclear and Coulomb components of $^7$Li breakup. For this purpose, we followed the procedures 
of Ref.~\cite{Otomar13} in their study of $^6$Li breakup. Fig.~\ref{fig4} shows plots of $\sigma_N$ versus $A_{\scr T}^{\scr 1/3}$.  
One observes that the nuclear components of the breakup cross section at a fixed value of $E/V_{\scr B}$ increase linearly with $A_{\scr T}^{\scr 1/3}$, to a 
good approximation.  On the other hand, Fig.~\ref{fig5} shows plots of $\sigma_C$ versus $Z_{\scr T}$. One notices that the cross sections increase
with $Z_{\scr T}$, showing a roughly linear behavior. These findings are analogous to those of Ref.~\cite{Otomar13}, for the $^6$Li Lithium isotope.

\section{Summary}

In summary, we have extended our investigation of the elastic breakup of weakly bound nuclei to a two-cluster projectile with significant dipole strength at low excitation energy.
The current work complements a previous one where no or very weak dipole strength is found. The isotopes of Lithium, $^7$Li, studied in the current paper, and $^6$Li are used for the purpose of comparison. We have found the same qualitative behavior in both cases, involving the Coulomb, nuclear and interference parts of the breakup cross section, namely,
a strong interference term and similar scaling laws for both the Coulomb and nuclear components of the breakup cross section, i.e.,
increasing linearly with $A_{\scr T}^{\scr 1/3}$ and $Z_{\scr T}$, respectively, for the same relative energy. The
comparison of $^7$Li with the $^{6}$Li elastic breakup shows that the $^{6}$Li total breakup
and its nuclear and Coulomb components are greater than for $^{7}$Li, for the same targets
and relative energies, whereas the ratios Coulomb/ nuclear components are
much larger for $^{7}$Li than for the corresponding $^{6}$Li system. We interpret those results in terms of the smaller breakup Q-value  in $^6$Li, and the low energy Coulomb dipole strengths of the Lithium isotopes. The results also indicate the importance of the Coulomb breakup through the excitation of higher multipolarities (quadrupole, octopole etc.) in the $\alpha$ +d cluster component of the $^6$Li wave function.

\noindent \textbf{Acknowledgements} \medskip \noindent We thank Pierre Descouvemont for useful comments. The authors acknowledge financial support from CNPq, CAPES, FAPERJ and FAPESP.

\end{document}